\documentclass[11pt]{article}

\usepackage[margin=1in]{geometry}
\usepackage{amsmath,amssymb,amsfonts}
\usepackage{booktabs}
\usepackage{array}
\usepackage{longtable}
\usepackage{graphicx}
\usepackage{float}
\usepackage{authblk}
\usepackage{setspace}
\usepackage{hyperref}
\usepackage{tabularx}
\usepackage[numbers,sort&compress]{natbib}

\hypersetup{
  colorlinks=true,
  linkcolor=blue,
  citecolor=blue,
  urlcolor=blue
}

\title{TRACE: Target-Trial Relevance Assessment and Compatibility-Guided Evidence Synthesis}

\author[1]{Wei Wei\thanks{Corresponding author: \texttt{wei.wei@yale.edu}}}
\author[2]{Liangze Ke}
\author[1]{Tianzhi Tang}
\author[1]{Pavan Challa}
\author[1]{Tulika Rudra Gupta}
\author[1]{Guannan Gong}
\author[3]{Satrajit Roychoudhury}
\author[1]{Raghav Sundar}

\affil[1]{Yale Cancer Center, Yale School of Medicine}
\affil[2]{Yale School of Public Health}
\affil[3]{Pfizer Inc}

\date{}

\begin{document}

\maketitle

\begin{abstract}
\textbf{Background:} External information is routinely used to inform oncology
trial design, but conventional meta-analysis often pools outcome summaries
without explicitly accounting for differences in eligibility criteria, trial
design, and enrolled populations. Such estimates may represent broad averages
over heterogeneous studies and may have limited relevance for the design of
future clinical trials.

\textbf{Methods:} We developed TRACE, a Target trial Relevance Assessment and
Compatibility-guided evidence synthesis framework using trial metadata. For
each external trial, structured eligibility criteria and baseline
characteristic summaries were extracted from published reports and
registries. An eligibility screen defined the candidate set, and each
candidate was assigned a posterior probability of compatibility with the target
trial, computed by Bayes' rule based on the largest standardized difference in
jointly reported baseline characteristics. Trials more likely compatible were retained,
and their reconstructed control-arm survival data were synthesized using a
Bayesian hierarchical model to generate a target-trial-aligned
external benchmark. We evaluated TRACE in a gastric/gastroesophageal junction
(G/GEJ) cancer case study and assessed its benchmarks by leave-one-trial-out
validation.

\textbf{Results:} We assembled eleven randomized phase III trials in first-line
HER2-negative advanced G/GEJ adenocarcinoma. With KEYNOTE-859 as the target
trial, three external trials (CheckMate-649, KEYNOTE-062 and LEAP-015) were
selected. The TRACE benchmark for control-arm median overall survival was 11.8
months (95\% posterior predictive interval, 8.9 to 15.2), against an observed
KEYNOTE-859 control-arm median of 11.8 months (95\% CI, 10.9 to 12.7). In
leave-one-trial-out validation, seven of the eleven trials had at least one
compatible external trial. TRACE's 95\% posterior predictive intervals contained the observed median in all seven, whereas the 95\% confidence intervals from KM estimates pooling all other trials' control arms contained it in five. Across these seven targets, mean pointwise curve coverage was 92\% for TRACE and 59\% for pooled KM. For the remaining four trials, no external trial was compatible, and TRACE returned a weakly informative benchmark, reflecting the absence of compatible external trials.

\textbf{Conclusions:} TRACE provides a transparent approach for selecting
external trials and synthesizing survival evidence according to target trial
relevance. By combining eligibility screening and baseline-characteristic
compatibility with reconstructed survival data, TRACE generates control-arm
benchmarks where compatible evidence exists and explicitly
weakly informative where it does not, supporting trial design and external
benchmark construction in oncology drug development.
\end{abstract}

\section{Introduction}

External trials provide important benchmarks for evaluating the efficacy of new treatments in oncology clinical trials. By establishing the expected outcomes under existing standards of care, external evidence helps investigators interpret emerging efficacy signals, formulate hypotheses regarding clinically meaningful treatment effects, and inform assumptions for sample size and power calculations. When patient accrual is challenging, external evidence may also be used to augment the internal control arm, potentially reducing sample size and trial costs. However, these applications have become increasingly challenging as oncology trials grow more complex and focus on molecularly defined patient subgroups. External trials available for a given clinical question may differ systematically from one another and from the trial being planned, particularly in eligibility criteria and patient characteristics. Consequently, simply aggregating all available external evidence may produce benchmarks that are poorly aligned with the intended target population. A more relevant approach is to synthesize external evidence according to its compatibility with the patient population and clinical context of the trial being designed.

To address this need, we propose TRACE (Target-trial Relevance Assessment and Compatibility-guided Evidence Synthesis), a framework for generating target-trial-aligned external benchmarks. In TRACE, the target trial is an existing or hypothetical trial that defines the patient population and clinical context for evidence synthesis through its eligibility criteria, baseline patient profile, treatment, comparator, and endpoint. It may be a completed trial whose population is of interest for clinical or regulatory decision-making, or a planned trial requiring external evidence to inform its design. This use is distinct from target trial emulation in the Hernán--Robins framework\cite{hernan2016using, hernan2022target}, which uses observational data to emulate a hypothetical randomized trial and estimate causal treatment effects. Rather than aggregating outcomes across external trials meeting broad inclusion criteria, TRACE identifies trials most relevant to the target population and synthesizes their evidence to generate an appropriate survival benchmark. The objective is not to perform target trial emulation or causal inference, but to provide a transparent and reproducible approach to evidence synthesis to inform trial design.

TRACE makes three methodological contributions. First, it characterizes trial relevance using multimodal metadata from trial publications and registries. Eligibility criteria are often reported as free text, baseline characteristics are presented in tables, and survival outcomes are reported through Kaplan--Meier (KM) curves and numbers at risk. Advances in AI and large language models have made it increasingly feasible to extract and structure these sources at scale. Existing tools demonstrate the feasibility of extracting design-relevant information from trial reports and reconstructing survival data from published KM curves\cite{gong2026learning, lang2025resolve}.

Second, TRACE integrates clinical expertise with information on both the intended and observed trial populations to assess the compatibility of external trials with the target trial. Expert-defined eligibility criteria are first used to identify candidate external trials sharing key clinical and design features with the target trial. Among these candidates, TRACE evaluates the similarity of the populations actually enrolled using reported baseline covariate summaries. A Bayesian framework combines prior beliefs about trial compatibility with the observed between-trial differences in baseline characteristics to derive an integrated compatibility score (ICS). This approach brings together three complementary sources of information including expert knowledge about trial comparability, eligibility criteria defining the intended patient population, and baseline characteristics describing the population actually enrolled. The resulting ICS provides a transparent criterion for identifying sufficiently compatible external trials for target-trial-aligned evidence synthesis.

Third, TRACE synthesizes reconstructed survival data from the target-trial-aligned evidence set using a Bayesian hierarchical model based on the beta--Stacy process \cite{walker1997beta,gong2026learning}. As a Bayesian nonparametric prior for distribution function, the beta--Stacy process provides flexible survival modeling without imposing restrictive parametric assumptions. The hierarchical structure enables information sharing across selected external trials while accommodating residual between-trial heterogeneity. The resulting external benchmark is represented by a posterior predictive distribution of the survival function, which accounts for uncertainty in both the underlying survival distribution and between-trial heterogeneity. Furthermore, TRACE approximates the synthesized evidence using a conjugate beta--Stacy process prior, allowing the information contributed by external trials to be quantified through its effective sample size and incorporated into subsequent analyses.

The remainder of this paper is organized as follows. Section \ref{sec:case} introduces the gastric and gastroesophageal junction cancer case study and the external trials included in the analysis. Section \ref{sec:met} presents the TRACE methodology, including eligibility screening, Bayesian compatibility assessment, hierarchical survival modeling, and construction of informative priors from synthesized evidence. Section \ref{sec:result} presents the case study results, evaluates the reliability of TRACE-derived external benchmarks through leave-one-trial-out validation, and illustrates their application to external information borrowing. Section \ref{sec:diss} discusses the findings, limitations, and directions for future research.

\section{Case study}\label{sec:case}
We evaluated TRACE using eleven randomized phase 3 trials of first-line advanced gastric and gastroesophageal junction (G/GEJ) adenocarcinoma~\cite{shitara2020keynote062,shah2021gamma1,kang2022attraction4,janjigian2021checkmate649,zaanan2025gastfox,rha2023keynote859,xu2023orient16,qiu2024rationale305,zhang2025gemstone303,shitara2025leap015,shen2025compassion15}. These trials were selected by a team of gastric oncologists based on shared key eligibility and design criteria, including histologically confirmed advanced or metastatic G/GEJ adenocarcinoma, HER2-negative disease, first-line systemic therapy, and a control arm receiving a platinum--fluoropyrimidine chemotherapy doublet. LEAD-ONC~\cite{gong2026learning} was used to extract baseline covariate summaries and reconstruct individual patient survival data (IPD) from published KM curves of overall survival (OS). The extracted baseline characteristics are summarized in Table~\ref{tab:covariates}.

\begin{figure}[!htbp]
    \centering
\includegraphics[width=1\textwidth]{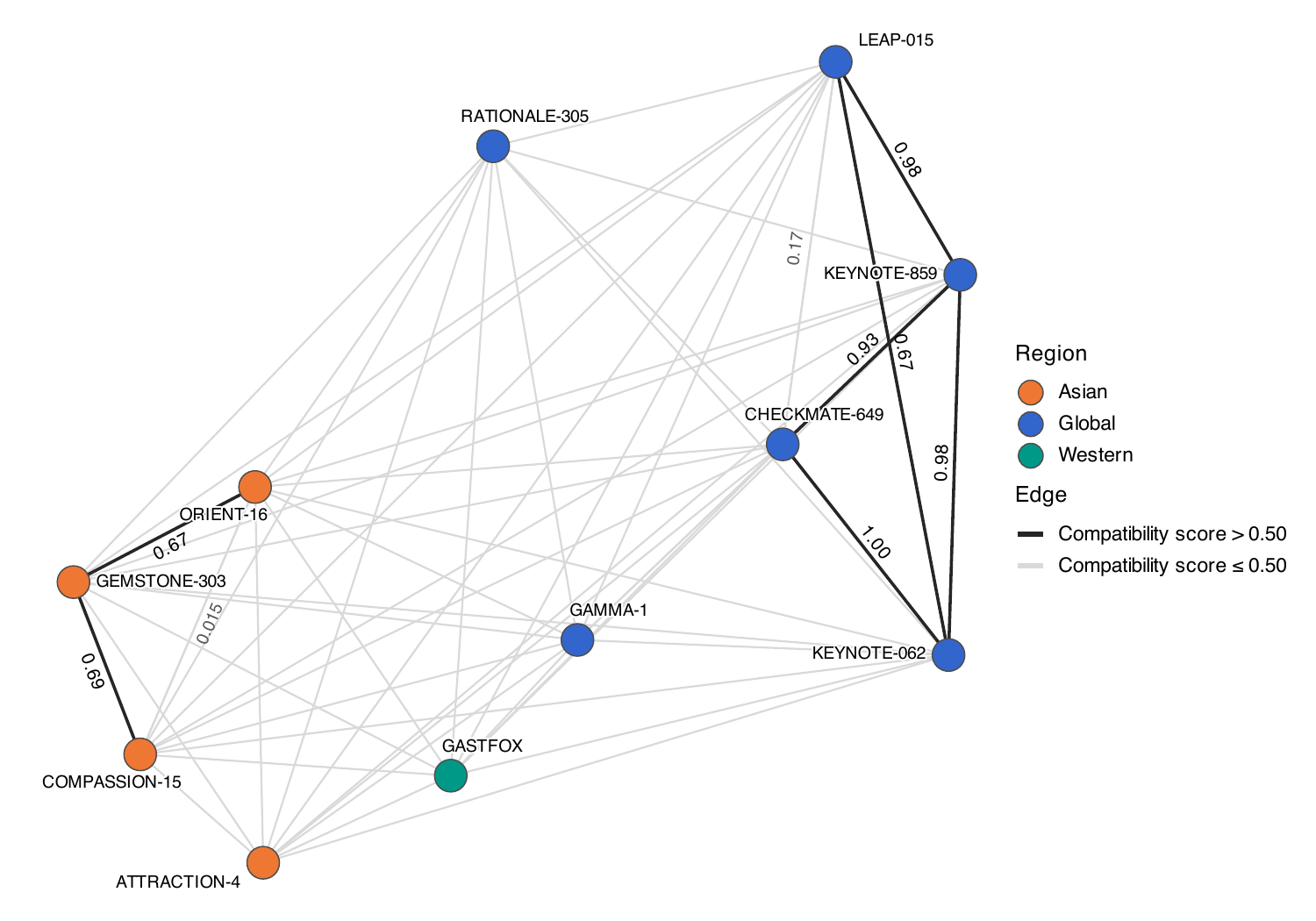}
    \caption{Trial compatibility network. Nodes represent trials. Edge labels show integrated compatibility scores.}
    \label{fig:network}
\end{figure}

For illustration purpose, we designated KEYNOTE-859 as a target trial and treated the remaining ten trials as candidate external trials for evidence synthesis. TRACE was employed to assess compatibility using the integrated compatibility score (ICS, Figure \ref{fig:network}) and synthesize reconstructed control-arm OS data from selected external trials. The resulting benchmark was compared with the observed KEYNOTE-859 control-arm OS (Figure~\ref{fig:post}).

\begin{figure}[!htbp]
    \centering
    \includegraphics[width=\textwidth]{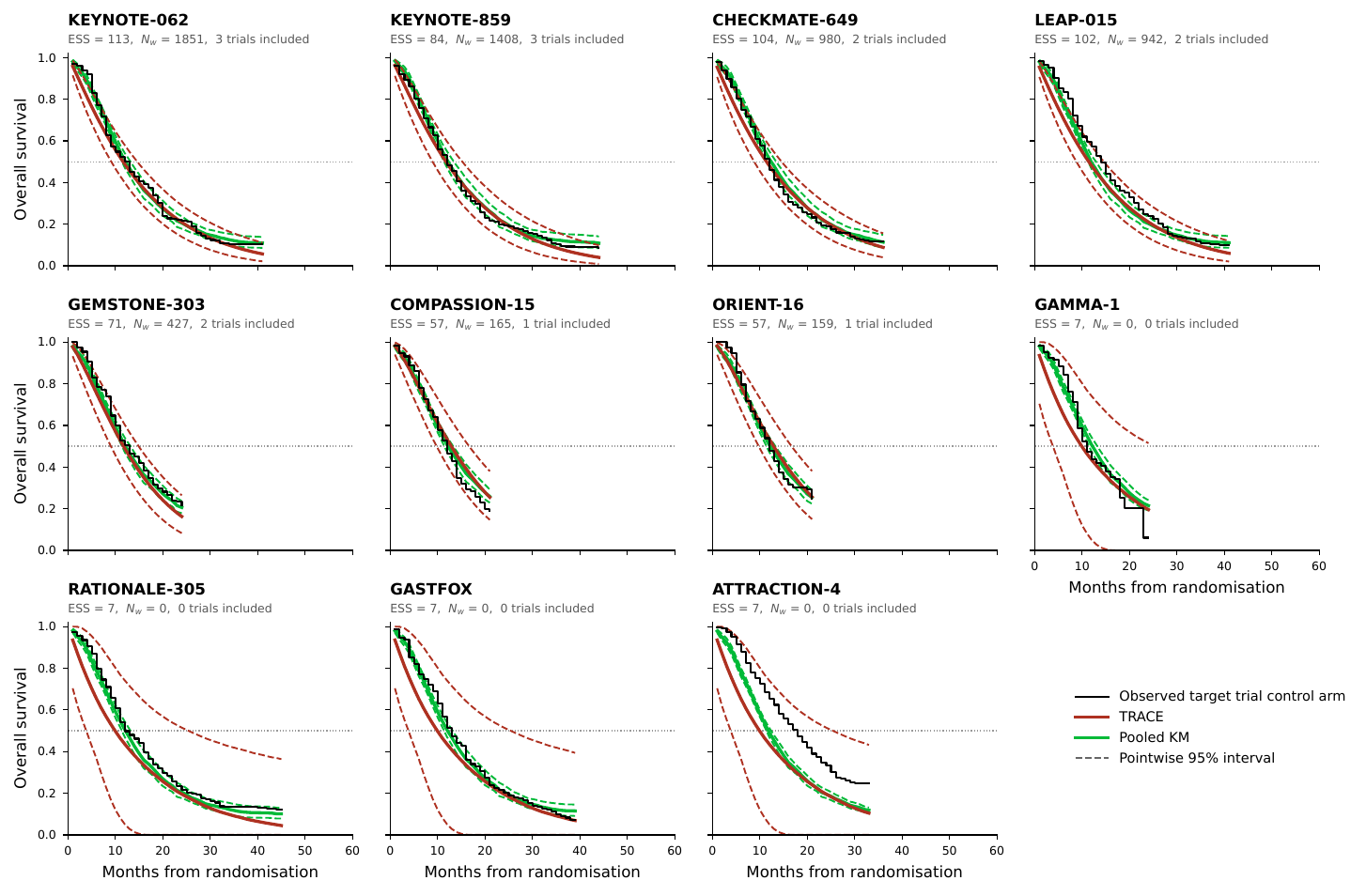}
    \caption{Leave-one-trial-out validation of target trial control-arm evidence synthesis. Each panel compares the held-out trial’s KM curve with the TRACE predictive benchmark and pooled KM estimate from all other trials. Predictive intervals and confidence limits are pointwise 95\%. Target trials without compatible external trials receive the prior predictive benchmark. }
    \label{fig:loo}
\end{figure}

We further conducted leave-one-trial-out validation across all eleven trials, comparing TRACE-derived external benchmarks with observed control-arm OS and benchmarks obtained by pooling all other trials without compatibility screening (Figure \ref{fig:loo}). We also examined the use of TRACE-derived benchmarks as informative priors to illustrate the potential benefits and risks of borrowing external information (Figure \ref{fig:forest}).

\section{Methods}\label{sec:met}
\subsection{Overview of TRACE}

Let $\mathcal{T}$ denote the target trial and $\mathcal{H}_1,\ldots,\mathcal{H}_J$ denote candidate external trials selected using expert-curated eligibility criteria, such as disease type and stage, line of therapy, biomarker requirements, and comparator treatment. These criteria define the shared clinical context and intended patient population of the candidate trials and $\mathcal{T}$. The baseline covariate profile of $\mathcal{T}$ is specified using reported patient characteristics for an existing trial or estimated from pilot studies or real-world data for a planned trial. TRACE then assesses compatibility between the $J$ external trials and the target trial by combining prior assumptions about compatibility with differences in their baseline covariate profiles, yielding an integrated compatibility score for each external trial. For a designated target trial $\mathcal{T}$, survival data from sufficiently compatible external trials are synthesized to generate a target-trial-aligned external benchmark.

\subsection{Trial metadata extraction}

For each candidate external trial $\mathcal{H}_j$, baseline covariate summaries were extracted from trial reports. IPD were reconstructed from the KM curves using Lead-Onc\cite{gong2026learning}, which provides automated metadata extraction, survival IPD reconstruction, quality checking of the reconstructed KM curve based on both visual and quantitative measures. 

TRACE does not use or generate patient-level baseline characteristics and does not support the development of predictive models relying on individual level predictors.

\subsection{Between-trial distance}
\label{sec:distance}

We quantify between-trial differences using standardized mean differences (SMDs) of baseline characteristics reported by both trials. For a continuous covariate $x_\ell$, the SMD between external trial $\mathcal{H}_j$ and target trial $\mathcal{T}$ is defined as \citep{austin2009balance}
\begin{equation}
d_{jT\ell} = \frac{\bar{x}_{j\ell}-\bar{x}_{T\ell}}
{\sqrt{(s_{j\ell}^{2}+s_{T\ell}^{2})/2}},
\label{eq:smd-cont}
\end{equation}
where $\bar{x}$ and $s$ denote the reported mean and standard deviation, respectively. 

When a continuous covariate is reported as a median with a range or interquartile range, we approximate the mean by the median and the standard deviation by the range divided by six or the interquartile range divided by $1.349$.

For characteristics reported as proportions, we clipped values to $[\varepsilon,1-\varepsilon]$, with $\varepsilon=0.05$, to avoid near-zero denominators. Let $p_{j\ell}$ and $p_{T\ell}$ denote the resulting proportions. The SMD is defined as
\begin{equation}
d_{jT\ell} = \frac{p_{j\ell}-p_{T\ell}}
{\sqrt{[p_{j\ell}(1-p_{j\ell})+p_{T\ell}(1-p_{T\ell})]/2}}.
\label{eq:smd-prop}
\end{equation}

Let $L_{jT}$ denote the number of baseline characteristics reported by both trials. We define the between-trial distance as the maximum absolute SMD:
\begin{equation}
D_{jT} = \max_{\ell=1,\ldots,L_{jT}} |d_{jT\ell}|.
\label{eq:dist}
\end{equation}

The maximum absolute SMD captures the largest observed discrepancy between two trials. Unlike an average distance measure, it prevents a substantial imbalance in one covariate from being offset by smaller differences in others. This conservative approach is particularly useful when compatibility assessment relies on published baseline summaries rather than individual patient-level covariate data.


\subsection{Integrated compatibility score}
\label{sec:selection}

TRACE assesses compatibility among trials that pass the expert-curated eligibility screen, ensuring that candidate external trials share key clinical and design features with the target trial. Within this candidate set, we use a Bayesian framework to evaluate compatibility based on the maximum absolute standardized mean difference, $D_{jT}$, defined in Section~\ref{sec:distance}.

Let $Z_j=1$ indicate that external trial $\mathcal{H}_j$ is compatible with target trial $\mathcal{T}$, and $Z_j=0$ otherwise. We specify two working reference models describing the distribution of SMDs under compatibility and incompatibility. For compatible trials, we assume the enrolled populations have similar baseline covariate profiles, so the signed SMDs are centered at zero:
\[
d_{jT\ell}\mid Z_j=1 \sim N(0,\sigma_1^2).
\]
We set $\sigma_1=0.10$, so that the probability of the absolute SMD for an individual characteristic is below 0.20 is approximately 95\%.

For incompatible trials, we allow systematic differences in baseline characteristics, represented by a nonzero mean shift:
\[
d_{jT\ell}\mid Z_j=0
\sim N(\pm\delta_0,\sigma_0^2).
\]
We set $\delta_0=0.25$, motivated by commonly used covariate balance-diagnostic guidelines \citep{stuart2010matching}. We specify $\sigma_0^2=10\sigma_1^2$ to allow substantially greater variation under incompatibility. These distributions serve as prespecified reference models for distinguishing compatible from incompatible baseline profiles, rather than exact sampling distributions of the observed SMDs.

Because compatibility depends on the magnitude rather than the direction of baseline differences, we model the absolute SMDs. Let $\Phi$ and $\phi$ denote the standard normal cumulative distribution function and probability density function, respectively. Under compatibility, $|d_{jT\ell}|$ follows a half-normal distribution with cumulative distribution function and density
\[
Q_1(x)=2\Phi\left(\frac{x}{\sigma_1}\right)-1,
\qquad
q_1(x)=\frac{2}{\sigma_1}\phi\left(\frac{x}{\sigma_1}\right),
\qquad x\geq 0.
\]

Under incompatibility, $|d_{jT\ell}|$ follows a folded-normal distribution:
\[
Q_0(x)=
\Phi\left(\frac{x-\delta_0}{\sigma_0}\right)
+\Phi\left(\frac{x+\delta_0}{\sigma_0}\right)-1,
\]
\[
q_0(x)=
\frac{1}{\sigma_0}
\left\{
\phi\left(\frac{x-\delta_0}{\sigma_0}\right)
+\phi\left(\frac{x+\delta_0}{\sigma_0}\right)
\right\},
\qquad x\geq 0.
\]
The folded-normal distribution is identical for positive and negative mean shifts of the same magnitude.

Because correlations among these characteristics cannot generally be recovered from published marginal summaries, we adopt a working assumption that their standardized differences are conditionally independent and identically distributed within each compatibility class. Under this assumption, the cumulative distribution function of the maximum absolute SMD is
\[
\Pr(D_{jT}\leq x\mid Z_j=z)
=
Q_z(x)^{L_{jT}},
\]
with corresponding density
\begin{equation}
f_{zj}(x)
=
L_{jT}Q_z(x)^{L_{jT}-1}q_z(x),
\qquad x>0.
\label{eq:f}
\end{equation}
Thus, the reference distribution of $D_{jT}$ explicitly accounts for the number of jointly reported baseline characteristics.

Let $\rho_j=\Pr(Z_j=1\mid\text{eligibility screen passed})$ denote the prior probability of compatibility before observing the baseline covariate differences. We set $\rho_j=0.5$ for all candidate trials, thereby assigning equal prior odds to the compatible and incompatible reference models. In practice, $\rho_j$ may differ across external trials if expert opinion favors certain trials. Applying Bayes' rule, we obtain
\begin{equation}
\pi_j
=
\Pr(Z_j=1\mid D_{jT},L_{jT},\text{screen passed})
=
\frac{\rho_j f_{1j}(D_{jT})}
{\rho_j f_{1j}(D_{jT})+(1-\rho_j)f_{0j}(D_{jT})}.
\label{eq:pi}
\end{equation}

We refer to $\pi_j$ as the integrated compatibility score (ICS). It integrates three sources of information, including expert-defined eligibility criteria that establish the candidate trial set, prior assumptions on compatibility, and observed differences in the baseline characteristics of the enrolled populations. The ICS represents a model-based posterior compatibility probability conditional on the prespecified reference models. Its numerical value should be interpreted as a compatibility screening measure rather than the precise probability of trial compatibility.

An important feature of the ICS is its adjustment for the number of jointly reported characteristics. Under compatibility, the maximum absolute SMD tends to increase as more characteristics are compared. Consequently, the same observed maximum can provide varying strength of evidence about compatibility depending on $L_{jT}$. For example, for $D_{jT}=0.276$, the ICS is 0.67 when five characteristics are compared. It increases to 0.80 and 0.89 when six and seven characteristics are compared, respectively. Thus, the assessment accounts for the number of comparisons underlying the observed maximum.

\subsection{Target-trial-aligned evidence synthesis}
\label{sec:synthesis}

External trials with low integrated compatibility scores may differ substantially from the target trial in patients actually enrolled. Conventional meta-analysis typically includes all eligible external trials regardless of their relevance to the target population, which may yield estimates that are overly broad or not directly applicable to the target trial.

We use ICSs as a ranking and screening measure to exclude external trials unlikely to be compatible with the target trial from evidence synthesis. Let
$$
\mathcal C=\{\mathcal H_j:\pi_j>\eta\}
$$
denote the target-trial-aligned evidence set, where $\eta\geq0$ is a prespecified compatibility threshold. All candidate external trials will be included if $\eta=0$.

\subsubsection{Review of the beta-Stacy process}

We model reconstructed survival outcomes using the beta--Stacy process (BSP), a Bayesian nonparametric prior for survival distributions,  which is conjugate to right-censored data and can be represented through independent beta-distributed discrete-time hazards\cite{walker1997beta}. 

Let $F(t)$ denote the cumulative distribution function (CDF) of discretized event times belonging
to a prespecified regularly spaced time grid 
$$
\boldsymbol{\tau}^*=\{t_1,t_2,\ldots\}.
$$
The grid may be specified in weeks or months depending on the time scale and duration of follow-up to minimize information loss. Event and censoring times are discretized by assigning them to the nearest grid point.

Let $S(t)=1-F(t)$ denote the survival function. 
Define the increments \[ \theta_k = F(t_k)-F(t_{k-1}), \qquad F(0)=0. \] Conditional on $\theta_{1}, \dots, \theta_{k-1}$, the random increment $\theta_{k}$ follows a beta-Stacy distribution \cite{mihram1967bivariate} with density function given by 
$$ f(\theta_{k}|\alpha_k,\beta_k, \theta_{1}, \dots, \theta_{k-1})=\frac{\Gamma(\alpha_k+\beta_k)}{\Gamma(\alpha_k)\Gamma(\beta_k)}\theta_{k}^{\alpha_k-1}\frac{(1-\theta_{1}-\cdots-\theta_{k})^{\beta_k-1}}{(1-\theta_{1}-\cdots-\theta_{k-1})^{\alpha_k+\beta_k-1}}, $$ 
where $\alpha_k >0$ and $\beta_k >0$. 

Define the hazard function as \[ h_k=\frac{\theta_k}{1-F(t_{k-1})}, \] 
which are distributed independently as 
\[ h_k \sim \text{Beta}(\alpha_k,\beta_k). \] 

We can construct the distribution of $F(t)$ based on $$F(t_k)=\sum_{j=1}^k \theta_j = 1-\prod_{j=1}^k (1-h_j).$$ Walker and Muliere\cite{walker1997beta} refer to the prior distribution of $F(t_k)$ as a discrete-time beta-Stacy process (BSP) with positive parameters $\{\alpha_k, \beta_k\}$ and jumps at the prespecified time grid $\boldsymbol{\tau}^*$. We denote this as 
\[ F\sim\operatorname{BSP}
  \{(\alpha_k,\beta_k):k\geq1\}.\]

\subsubsection{Hierarchical Bayesian survival model}

Let $F_j$ denote the CDF for an external trial in the target-trial-aligned evidence set ($\mathcal H_j\in\mathcal C$), and let $F^*$ denote the CDF for the target trial. We assume

$$
F_j,F^* \sim \operatorname{BSP}
  \{(\alpha_k,\beta_k):k\geq1\}.
$$

Let $h_{jk}$ and $h^*_{k}$ denote the discrete-time hazard for $\mathcal H_j$ and the target trial. Under the BSP representation,
$$
h_{jk}, h^*_{k}\sim \operatorname{Beta}(\alpha_k,\beta_k).
$$
The reconstructed survival data contribute through the binomial likelihood
$$
d_{jk}\sim \operatorname{Binomial}(r_{jk},h_{jk}),
$$
where $d_{jk}$ and $r_{jk}$ denote the number of events and the number of patients at risk for $\mathcal H_j$ at $t_k$.

Following Walker and Muliere \cite{walker1997beta}, the BSP prior can be parameterized by a centering function $G(t)$ and a precision function $c(t)$, such that
$$
\alpha_k=c(t)\{G(t_{k-1})-G(t_k)\},\qquad
\beta_k=c(t)G(t_k).
$$

Under this parameterization, the prior mean survival function is

$$
E\{S(t)\}=G(t),
$$
so $G(t)$ represents the centering survival curve, while $c(t)$ controls the strength of prior information around this curve. 

Motivated by empirical evidence that survival outcomes in oncology trials are often well described by two-parameter Weibull models \cite{plana2022cancer}, we specify
$$
G(t)=\exp(-\lambda t^\kappa),
$$
where $\lambda>0$ and $\kappa>0$ are scale and shape parameters. 

Although we focus on a Weibull model in this work, more flexible choices such as the piecewise exponential model can also be considered for $G(t)$.

This hierarchical model allows each $\mathcal H_j$ to have its own hazard function modeled as exchangeable draws from a common $\operatorname{Beta}(\alpha_k,\beta_k)$, whose parameters are governed by shared hyperparameters $(\lambda,\kappa,c(t_k))$. Trial-specific survival curves are therefore shrunk toward the centering survival curve $G(t)$, and the degree of shrinkage is controlled by the concentration parameter $c(t_k)$. 

We assume $c(t)=c$, so a single parameter governs between-trial heterogeneity across the entire follow-up period. This is a necessary choice in most applications. As the number of selected trials is typically small, data cannot support reliable estimation of time-specific concentration parameters. 

A small $c$ allows the trial-specific hazards to depart substantially from $G(t)$, accommodating large between-trial heterogeneity, whereas a large $c$ pulls them tightly toward $G(t)$, approaching complete pooling. Because $c$ is assigned a prior, the strength of borrowing is learned from the external data and accounts for between-trial heterogeneity.

Using the hierarchical BSP model described above, the survival distribution for a new trial population is generated from 
\[ S^*(t)=\prod_{t_k\le t}(1-h_k^*). \] 
Posterior predictive draws of $h_k^*$ therefore generate a distribution of possible survival curves $S^*(t)$. These draws characterize uncertainty in survival outcomes for the target population and form the basis for projected survival experiences.

\subsection{Prior choice}

Under the BSP parameterization, $\alpha_1+\beta_1=c$, so the concentration parameter $c$ can be interpreted as the number of pseudo-patients initially at risk. We assume $c$ is unknown and will estimate $c$ by assigning it a prior. To relate the prior concentration to the amount and compatibility of external evidence, we define the compatibility-weighted sample size as
\[
N_w=\sum_{\mathcal H_j\in\mathcal C}\pi_j n_j,
\]
where $n_j$ is the sample size of the external arm contributing survival data. We assign
\[
c\sim\operatorname{Uniform}(c_0,\max(N_w,c_0)).
\]
This specification ensures $c_0\leq c\leq N_w\leq N$ when
$N_w\geq c_0$, where $N=\sum_{\mathcal H_j\in\mathcal C}n_j$. The upper bound allows greater prior concentration when more compatible external evidence is available, while the lower bound $c_0$ prevents excessive departures from the centering distribution $G(t)$. When no external trials are retained or the compatibility-weighted sample size is below $c_0$, we set $c=c_0$.

We specify a Weibull centering survival function
\[
G(t)=\exp\left\{-\log(2)\left(\frac{t}{m}\right)^\kappa\right\},
\]
where $m$ and $\kappa$ denote the median survival time and shape parameter, respectively. We assign independent priors
\[
\log m\sim N(\log 10,0.40^2),\qquad
\log\kappa\sim N(0,0.5^2).
\]
The prior median of $m$ is 10 months, with a 95\% interval of 4.6--21.9 months. This choice is informed by randomized trials of first-line chemotherapy for advanced gastric cancer, which reported average median overall survival of 8.6 months in trials published through 2005 and 10.6 months in later trials \citep{kawakami2013postprogression}, and a systematic review reporting median overall survival of 5.0--13.1 months across treatment arms in Western populations \citep{marupuru2023firstline}. The prior accommodates these external outcomes while allowing substantial uncertainty. The prior on $\kappa$ has a median of 1, corresponding to a constant hazard, while allowing departures from this assumption.

We set the lower bound $c_0=50$, which is fixed throughout the analysis. When no compatible external trial is retained, we have $N_w=0, c=c_0$, so no estimation of $c$ is required. The resulting prior predictive survival distribution nevertheless incorporates uncertainty in $G$, yielding a broad predictive distribution for median survival centered around 10 months, with a wider 95\% interval of approximately 3.9--25.3 months. As described in Section~\ref{sec:approx}, approximating this prior predictive distribution with a single BSP with fixed $G(t)$ and $c$ through mean and variance matching yields $\widetilde c\approx7$, corresponding to approximately seven pseudo-patients initially at risk. This effective concentration is substantially smaller than $c_0$ because it accounts for uncertainty in $G$. 

\subsection{Approximation of the predictive prior}
\label{sec:approx}

The hierarchical BSP model assumes trial-specific survival functions are drawn from a common BSP with centering distribution $G$ and precision parameter $c$. The posterior predictive distribution for a new target-trial-aligned control arm incorporates uncertainty in both $G$ and $c$, resulting in a distribution that cannot be represented by a BSP with fixed parameters. We approximate this distribution using a BSP with fixed Weibull centering function $\widetilde G(t)$ and precision $\widetilde c$, chosen to match its predictive mean and variance. The resulting approximated BSP is conjugate to right-censored survival data and can be used as an informative prior in subsequent analyses.

Let $\{S_{\mathrm{new}}^{(b)}(t_k):b=1,\ldots,B;\ k=1,\ldots,K\}$ denote $B$ posterior predictive draws of the survival function on the prespecified time grid, where $t_K$ is the largest follow up time among selected external trials or a prespecified  time horizon. We estimate the pointwise predictive mean and variance as
\begin{equation}
\mu_k=\frac{1}{B}\sum_{b=1}^{B}S_{\mathrm{new}}^{(b)}(t_k),
\qquad
v_k=\frac{1}{B-1}\sum_{b=1}^{B}
\{S_{\mathrm{new}}^{(b)}(t_k)-\mu_k\}^2.
\label{eq:predsummary}
\end{equation}

For a BSP with fixed $c$ and $G(t)$, the pointwise mean and variance of the survival function at $t_k$ have closed-form expressions:
\begin{equation}
E\{S(t_k)\mid G,c\}=G(t_k),
\qquad
\operatorname{Var}\{S(t_k)\mid G,c\}
=\frac{G(t_k)\{1-G(t_k)\}}{c+1}.
\label{eq:bspmoments}
\end{equation}

We use these expressions to solve $c$ and $G(t)$ in two steps. First, we fit a Weibull survival function to the predictive mean by minimizing the residual sum of squares:
\begin{equation}
(\widetilde\kappa,\widetilde\lambda)
=\arg\min_{\kappa,\lambda>0}
\sum_{k=1}^{K}
\{\exp(-\lambda t_k^\kappa)-\mu_k\}^2.
\label{eq:step1}
\end{equation}
This yields the approximating centering function
$\widetilde G(t)=\exp(-\widetilde\lambda t^{\widetilde\kappa})$.

Likewise, we solve $c$ in the second step
\begin{equation}
\widetilde c=\arg\min_{c>0}
\sum_{k=1}^{K}
\left\{
\frac{\widetilde G(t_k)\{1-\widetilde G(t_k)\}}{c+1}
-v_k
\right\}^2.
\label{eq:step2}
\end{equation}

The resulting approximated BSP has independent beta-distributed hazards with parameters
\[
\widetilde\alpha_k=
\widetilde c\{\widetilde G(t_{k-1})-\widetilde G(t_k)\},
\qquad
\widetilde\beta_k=
\widetilde c\,\widetilde G(t_k).
\]
These parameters have a pseudo-count interpretation: $\widetilde\alpha_k$ represents the prior number of events at $t_k$, while $\widetilde\beta_k$ represents the number of pseudo-patients surviving beyond $t_k$. Their sum,
\[
\widetilde\alpha_k+\widetilde\beta_k
=\widetilde c\,\widetilde G(t_{k-1}),
\]
represents the number of pseudo-patients at risk immediately before $t_k$. 

In particular, $\widetilde c$ represents the effective sample size (ESS) for the approximated predictive distribution at $t_0$. We further define the effective number of events (ENE) over the follow-up period as
\begin{equation}
\widetilde{\mathrm{ENE}}
=\sum_{k=1}^{K}\widetilde\alpha_k.
\label{eq:ene}
\end{equation}
Together, $\widetilde c$ and $\widetilde{\mathrm{ENE}}$ quantify the information in the synthesized external benchmark in terms of equivalent initial pseudo-patients and expected pseudo-events through $t_K$.

When no compatible external trial is retained, the same approximation is applied to the prior predictive distribution.

\subsection{Leave-one-trial-out validation}

We evaluated the ability of TRACE to generate reliable external benchmarks using leave-one-trial-out validation. In each iteration, one trial was held-out and treated as the target trial,  while the remaining trials were used as candidate external trials for evidence synthesis. TRACE was then applied to compute ICSs relative to the target trial. Trials exceeding the prespecified threshold $\eta$ formed the target-trial-aligned evidence set for that iteration.
The Bayesian hierarchical model was fitted to reconstructed external control IPD from the selected evidence set, and the resulting predictive benchmark was compared with the observed control-arm survival experience of the target trial. Agreement was assessed by whether the observed median OS of the target trial fell within the 95\% posterior predictive interval for the benchmark median and whether the observed KM curve was largely contained within the pointwise predictive bands.

To evaluate TRACE against conventional practice, we compared the predictive benchmark with KM estimators obtained by pooling all external controls without compatibility-based selection. We treated trials as clusters and obtained 95\% intervals using robust standard errors.

To illustrate the risks and benefits of borrowing external information, we
performed an additional analysis treating the TRACE predictive benchmark as an
informative prior for the held-out trial's own control arm. For each held-out
trial $\mathcal T$, the prior was the beta-Stacy process with fixed
$\widetilde G$ and $\widetilde c$ that approximates the predictive distribution from TRACE (Section~\ref{sec:approx}). Let $d_{kT}$ and
$r_{kT}$ denote the number of events and the number of patients at risk at
$t_k$ in the reconstructed IPD for the control arm of $\mathcal T$. Because the beta-Stacy process
is conjugate for right-censored data, the posterior is again a beta-Stacy
process, with independent hazards
\[
h_k \mid d_{kT}, r_{kT} \sim \mathrm{Beta}\bigl(\widetilde a_k + d_{kT},\;
\widetilde b_k + r_{kT} - d_{kT}\bigr), \qquad k=1,\dots,K .
\]
Posterior summaries of the median are based on 20,000 draws from the posterior distribution of survival curves.

This analysis is intended not as a validation of TRACE for information borrowing but as a demonstration of the bias--variance trade-off of borrowing, particularly when the held-out trial differs from the trials judged compatible
with it.

\section{Results}\label{sec:result}
Throughout this section, the TRACE benchmark for a target trial is the posterior
predictive distribution of the Bayesian hierarchical model in
Section~\ref{sec:synthesis}, or the prior predictive distribution when no external
trial is compatible with the target. Predictive summaries of benchmark in
Figures~\ref{fig:post}, ~\ref{fig:loo} and ~\ref{fig:forest} are all computed using 20,000 draws from this distribution. The beta-Stacy approximation of Section~\ref{sec:approx}
is used only to estimate
ESS and ENE, and to function as the informative
prior when borrowing is considered.

\subsection{Compatibility assessment}
All eleven trials reported performance status, geographic region, primary tumor location, age, and sex. Eight trials reported the prevalence of liver metastases, and seven reported the prevalence of peritoneal metastases. For each pair of trials, compatibility was assessed using only the baseline characteristics reported by both. We set the compatibility threshold at $\eta=0.50$, retaining trials that were more likely to be compatible with the target trial.

Figure~\ref{fig:network} shows the integrated compatibility scores (ICSs). Of the 55 trial pairs, seven had ICSs above $0.50$ and 46 below $0.001$. The remaining two pairs were CHECKMATE-649–LEAP-015 ($D_{jT}=0.33$, $\pi_j=0.17$) and COMPASSION-15–ORIENT-16 ($D_{jT}=0.41$, $\pi_j=0.015$). The compatible pairs formed two groups, with the grouping largely reflecting differences in the proportion of patients enrolled in Asia.

The first group comprised four global trials
enrolling 22\% to 37\% of patients in Asia, including CheckMate-649, KEYNOTE-062, KEYNOTE-859 and
LEAP-015. Only the CheckMate-649--LEAP-015
pair fell below the compatibility threshold of 0.50. The other trial pairs in this group have an ICS of at least $0.67$ and maximum standardized differences between $0.14$ and $0.28$. 

The second group comprised three of the four all-Asian trials: GEMSTONE-303, COMPASSION-15, and ORIENT-16. GEMSTONE-303 was compatible with both other trials, whereas COMPASSION-15 and ORIENT-16 were not compatible with each other. 

No all-Asian trial was compatible with any trial enrolling outside Asia. The smallest maximum standardized difference across these comparisons was \(0.58\), observed for RATIONALE-305 with either ORIENT-16 or ATTRACTION-4. For comparisons with the remaining trials enrolling outside Asia, the minimum was approximately \(1.55\). Four trials had no compatible partner. ATTRACTION-4 differed from the closest all-Asian trial,
ORIENT-16, mainly in the prevalence of peritoneal metastases and performance status
($D_{jT}=0.57$). RATIONALE-305, with 75\% of patients enrolled in Asia, was intermediate between the two groups but compatible with neither. Its closest trials overall were ORIENT-16 and ATTRACTION-4 (\(D_{jT}=0.58\) for both), whereas its closest trial within the four-trial global group was LEAP-015 (\(D_{jT}=0.83\)). GAMMA-1 and GASTFOX, the two trials without Asian enrollment, differed from
each other mainly in the proportion with gastro-oesophageal junction or oesophageal
primary tumours ($D_{jT}=0.50$).

\subsection{Case study}

\begin{figure}[!htbp]
    \centering
\includegraphics[width=1\textwidth]{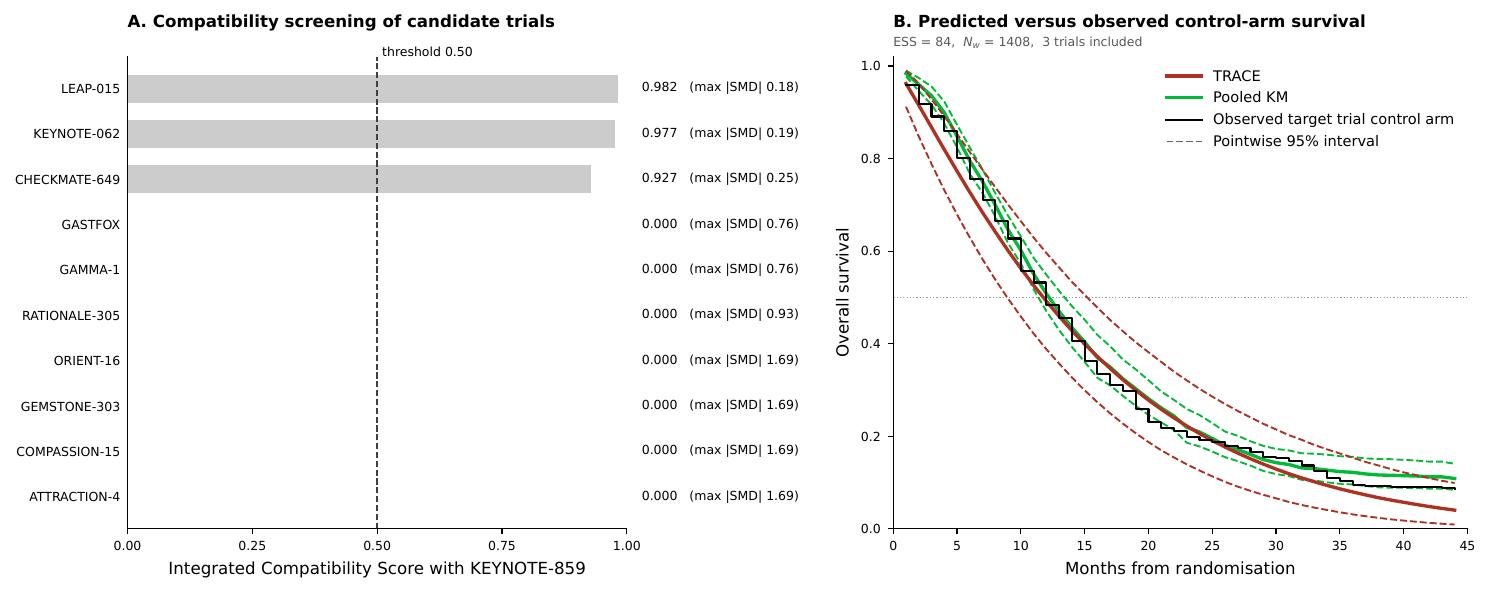}
    \caption{Case study: projecting the OS for the control arm of KEYNOTE-859, treating KEYNOTE-859 as the target trial. (A) Compatibility scores for the ten candidate external trials. (B) TRACE posterior predictive mean and pointwise 95\% predictive interval, compared with the observed target-trial KM curve and the KM estimate pooling all ten external control arms.}
    \label{fig:post}
\end{figure}

In this illustrative case study, KEYNOTE-859 served as the target trial. Its baseline characteristics were used to assess compatibility with the remaining ten trials, while its control-arm OS was held out from evidence synthesis. Three trials exceeded the prespecified compatibility threshold of $\eta=0.50$, including LEAP-015, KEYNOTE-062, and CheckMate-649 (Figure~\ref{fig:post}A). All three were global trials, with maximum absolute standardized differences from KEYNOTE-859 ranging from $0.18$ to $0.25$. The remaining seven trials had compatibility scores below $0.001$.

We fitted the BHM to reconstructed external control IPD from the three selected trials, whose compatibility-weighted sample size was $N_w=1408$. The resulting posterior predictive distribution provided a benchmark for the control arm of KEYNOTE-859 (Figure~\ref{fig:post}B). Its BSP approximation had an ESS of $84$. The predicted control-arm median OS was $11.8$ months ($95\%$ posterior predictive interval, $8.9$--$15.2$ months), in close agreement with the observed median OS for the control arm of KEYNOTE-859 (11.8, $95\%$ CI, $10.9$--$12.7$ months; $n=789$). The observed KM curve remained within the pointwise $95\%$ posterior predictive band throughout follow-up. For comparison, pooling the control-arm data from all ten other trials yielded a KM median of $12.2$ months ($95\%$ CI, $11.4$--$13.3$ months).

\subsection{Leave-one-trial-out validation}

\begin{figure}[!htbp]
    \centering
    \includegraphics[width=\textwidth]{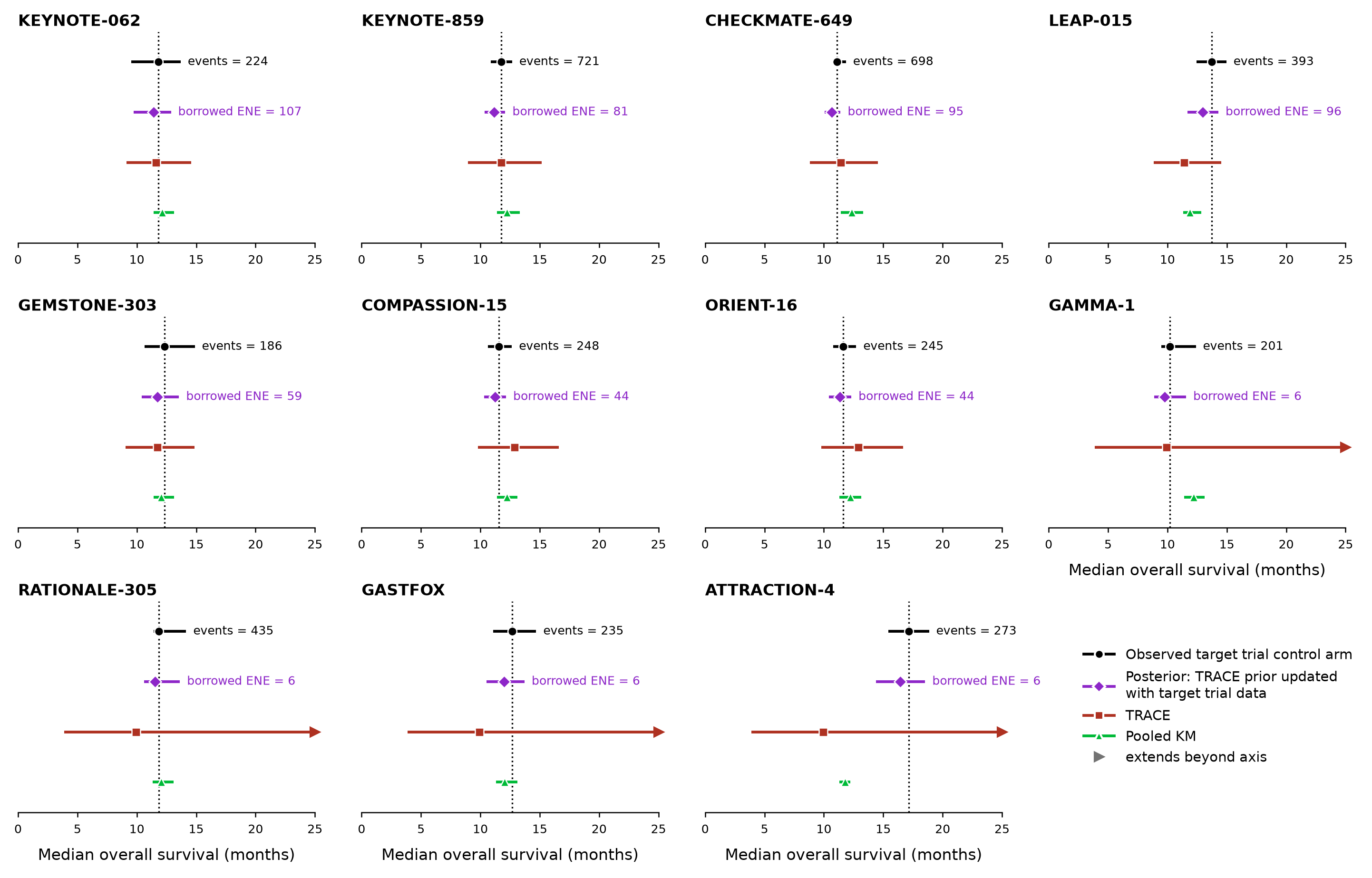}
    \caption{Leave-one-trial-out validation of target trial control-arm median OS estimates. Each panel corresponds to one trial designated as the held-out target trial. In each panel, median OS for the control arm of the target trial is estimated using the KM method, and compared to the TRACE predictive benchmark, posterior estimate after updating the TRACE prior with target-trial data, and pooled KM estimates from all other external trials. ENE denotes the effective number of event of the prior. }
    \label{fig:forest}
\end{figure}

Each of the eleven trials was in turn treated as the target
(Figures~\ref{fig:loo} and~\ref{fig:forest}). Seven targets had at least one compatible external trial: the four multiregional trials, each with two or
three trials in its evidence set ($N_w$ from $942$ to $1851$), and the three all-Asian trials GEMSTONE-303 (two trials, $N_w=427$), COMPASSION-15 and
ORIENT-16 (one trial each, $N_w=165$ and $159$). 

On these seven folds, the observed median OS fell within the
$95\%$ posterior predictive interval in every case. The mean interval width was $6.1$ months (range $5.4$ to $6.9$). LEAP-015 had the longest observed median ($13.7$ months) and TRACE predicted $11.4$ months ($8.8$ to $14.5$). In comparison, the pooled KM estimate from all other trials have narrower confidence intervals (mean width $1.8$ months) and covered the observed median in five of the same seven folds, missing CheckMate-649 and LEAP-015. 

To assess agreement over the whole survival curve, we evaluated the observed KM curve of each held-out target on a monthly grid $t_k=1,2,\dots$ up to the target's last follow-up time, and recorded the proportion of grid
points at which it lay within the pointwise $95\%$ posterior predictive interval of $S(t_k)$ for TRACE, or within the pointwise $95\%$ CI of the pooled KM estimate. We refer to this proportion as the pointwise curve coverage, and report its mean across folds. On the seven
folds with a compatible evidence set, the mean pointwise coverage was $0.92$ for TRACE, with low coverage for LEAP-015 (0.80) and ORIENT-16 (0.81). For the pooled KM estimator, the mean pointwise coverage was $0.59$, ranging from $0.47$ for CheckMate-649 to $0.79$ for GEMSTONE-303.

For the four targets without a compatible trial, ATTRACTION-4, GAMMA-1, GASTFOX and RATIONALE-305, the benchmark is the prior predictive distribution, which has a median of $10.0$ months ($95\%$ interval $3.9$ to $25.3$), and an effective sample size of approximately $7$. The width of these predictive intervals signals that no
compatible external evidence was available. 

Across all eleven folds, the
pooled Kaplan--Meier estimate covered seven, missing ATTRACTION-4, CheckMate-649, GAMMA-1 and LEAP-015.

Finally, we examined the consequences of using the TRACE benchmark as an
informative prior for each held-out trial's own control arm
(Figure~\ref{fig:forest}). For the seven trials with a compatible evidence set,
this prior contributed between $44$ (COMPASSION-15 and ORIENT-16) and $107$
(KEYNOTE-062) effective events, equivalent to $11\%$ to $48\%$ of the events observed in the held-out control arm. For the four trials without a compatible
external trial, this prior contributes about $6$ events. 

The benefit of
borrowing was a gain in precision where the borrowed evidence was large
relative to the held-out arm. The risk of borrowing was bias toward the external prior. It was largest for
LEAP-015, whose observed median of $13.7$ months was the longest among the
trials with compatible evidence. Borrowing $96$ events pulled the posterior
median down by $0.5$ months. For the other six trials the pull was at most
$0.22$ months. ATTRACTION-4 illustrates the
protective role of screening. Because no external trial was
compatible with it, the prior contributed only $6$ events and moved the posterior median by $0.14$ months. 

Because borrowed evidence was large only where compatibility was high, TRACE allows a more targeted borrowing approach by deciding which trial to borrow and how much information to borrow according to the compatibility of external trials with the target trial. Compatibility-based
selection combined with hierarchical modeling  delivered
target-trial-relevant external benchmarks whose precision reflected the amount of compatible evidence. 

\section{Discussion}\label{sec:diss}
By combining eligibility screening, baseline-characteristic compatibility assessment, and hierarchical survival modeling, TRACE provides a transparent framework for generating external survival benchmarks aligned with a target trial's population and clinical context, while accounting for residual heterogeneity among selected trials. Although meta-regression can help explain between-trial heterogeneity through trial-level covariates, its usefulness for predicting outcomes in a planned trial may be limited when eligibility screening leaves few candidate studies. As trial is the unit of analysis, estimating multiple covariate effects and residual heterogeneity can be unstable, particularly when baseline characteristics are incompletely reported. 

Several limitations warrant consideration. TRACE relies on the availability and quality of published trial information, and important prognostic factors may be missing or reported inconsistently. The ICS should therefore be interpreted as a model-based ranking and screening measure whose values depend on the available characteristics and prespecified reference models. Moreover, the target patient profile is treated as fixed. When this profile is unknown at the design stage, an anticipated profile may be estimated from electronic health record data or prior pilot studies by applying the planned trial's eligibility criteria. Because this profile may differ from that of the patients ultimately enrolled, analyses using alternative plausible profiles can help assess the sensitivity of the benchmark to the anticipated population composition. Finally, TRACE is designed for predictive evidence synthesis and trial planning. It is not a tool for target-trial emulation or causal inference. Target-trial alignment refers to the relevance of external evidence to the planned trial, assessed through eligibility criteria and reported baseline characteristics. Similarity in reported baseline summaries does not establish causal exchangeability or adjust for patient-level confounding. The resulting survival benchmark therefore represents a model-based prediction informed by selected external trials, without establishing a causal interpretation. 

In Future work, we plan to extend TRACE in several directions. Although the present application focuses on survival outcomes, TRACE could be extended to binary, continuous, and other types of outcomes by integrating
its eligibility screening and compatibility assessment with the meta-analytic-predictive prior to generate predictive benchmarks for the outcomes of interest in the target trial\cite{neuenschwander2010summarizing}, such as the probability of response. In addition, feature-specific weights or expert-informed priors could reflect the differing prognostic importance of individual baseline covariate features. Broader validation across tumor types, treatment settings, and endpoints will be important to characterize when target-trial-aligned external benchmarks are most reliable.

\newpage
\bibliographystyle{unsrtnat}
\bibliography{ref}

\newpage
\begin{table}[ht]
\centering
\caption{Baseline characteristics from the control arms of the eleven trials. Each pair of trials is compared on the
characteristics both report (NR, not reported). 
For GAMMA-1, the proportion enrolled in Asia is taken from the registered site list (NCT02545504), which includes no sites in East Asia. }
\label{tab:covariates}
\small
\setlength{\tabcolsep}{4pt}
\resizebox{\textwidth}{!}{%
\begin{tabular}{c|c|c|c|c|c|c|c|c}
  \hline
  Trial & $n$ & Age & ECOG $\geq 1$ & Asia & Male & GEJ/oesophageal & Liver & Peritoneal \\
        &     & mean (SD)               & (\%)          & (\%) & (\%) & primary (\%)   & metastases (\%) & metastases (\%) \\
  \hline
  COMPASSION-15 & 305 & 64.0 (8.2) & 76.4 & 100.0 & 77.0 & 24.3 & 46.9 & 12.5 \\
  GEMSTONE-303 & 238 & 63.0 (8.2) & 76.5 & 100.0 & 74.8 & 12.6 & 36.1 & 12.2 \\
  ORIENT-16 & 323 & 60.0 (11.1) & 71.8 & 100.0 & 71.2 & 18.6 & 39.6 & 19.5 \\
  ATTRACTION-4 & 362 & 65.0 (10.3) & 46.0 & 100.0 & 75.0 & 9.1 & 36.5 & 45.0 \\
  KEYNOTE-859 & 789 & 62.0 (12.6) & 62.0 & 33.0 & 69.0 & 23.4 & 39.4 & NR \\
  LEAP-015 & 437 & 61.0 (10.0) & 53.0 & 37.0 & 70.0 & 26.3 & NR & NR \\
  RATIONALE-305 & 496 & 61.0 (10.4) & 69.0 & 75.0 & 70.0 & 20.2 & 37.9 & 43.1 \\
  CHECKMATE-649 & 792 & 61.0 (11.1) & 57.0 & 22.0 & 71.0 & 29.8 & 39.6 & 23.7 \\
  GAMMA-1 & 214 & 63.0 (9.7) & 57.0 & 0.0 & 72.0 & 33.2 & NR & NR \\
  KEYNOTE-062 & 250 & 62.5 (10.7) & 54.0 & 24.4 & 71.6 & 26.8 & NR & NR \\
  GASTFOX & 253 & 63.9 (10.5) & 57.0 & 0.0 & 77.0 & 57.3 & 47.8 & 27.7 \\
  \hline
\end{tabular}%
}
\end{table}

\newpage

\newpage

\newpage

\newpage

\end{document}